\documentclass[final,5p,times,twocolumn,sort&compress]{elsarticle}
\usepackage{graphicx}
\usepackage{amsmath}

\begin{document}

\begin{frontmatter}

\title{Towards detailed tomography of high energy heavy-ion collisions by $\gamma$-jet}

\author{Guo-Liang Ma}
            
\address{Shanghai Institute of Applied Physics, Chinese
Academy of Sciences, P.O. Box 800-204, Shanghai 201800, China}


\begin{abstract}

Within a multi-phase transport (AMPT) model with string melting scenario,  the transverse momentum  imbalance between prompt photon and jet is studied in Pb+Pb collisions at $\sqrt{s_{_{\rm NN}}}$ = 2.76 TeV. Jet loses more energy in more central collisions due to strong partonic interactions between jet parton shower and partonic matter, which is more significant than due to hadronic interactions only. The hadronization and final-state hadronic interactions have little influences on the imbalance. The imbalance ratio $x_{j\gamma}$ is sensitive to both production position and passing direction of $\gamma$-jet, which provides an opportunity to do detail $\gamma$-jet tomography on the formed partonic matter by selecting different $x_{j\gamma}$ ranges. It is also proposed that $\gamma$-hadron azimuthal correlation associated with photon+jet is a probe to see the medium responses to different $\gamma$-jet production configurations. 

\vspace{1pc}
\end{abstract}


\end{frontmatter}

\section{Introduction}
\label{sec:intro}

Measurements of jets produced in hard scattering processes serve as an important probe of the strongly interacting partonic matter at RHIC and LHC which can help investigate the properties of the formed new matter~\cite{Appel:1985dq, Blaizot:1986ma}.  Many experimental observables show that jets lose their energies significantly because they have to interact with the hot and dense medium when they pass it through~\cite{Adams:2005dq, Adcox:2004mh}. Recent experimental results based on full jet reconstruction disclose more detailed characterizations of jet-medium interactions~\cite{Aad:2010bu, Chatrchyan:2011sx}. Prompt photon and jet, i.e. $\gamma$-jet, can be produced by a initial hard scattering process with similar large transverse momentum ($p_{T}$) back-to-back-ly at leading order. Recent results from PHENIX show that direct photons with high $p_{T}$ do not flow~\cite{Adare:2011zr} and their nuclear modification factor $R_{AA}$ is around unity ~\cite{Afanasiev:2012dg} , which is consistent with the picture that they are dominantly produced in initial hard scatterings and do not participate in strong interactions due to their neutral electric and color charges. $\gamma$-triggered correlation has been proposed as a golden channel for jet physics, since it brings a different kinematical and geometrical bias in comparison with jet-triggered one~\cite{Zhang:2009rn}, although it costs all additional information from the photon trigger. Recently, lots of experimental efforts have been invested in measuring reconstructed jets in high energy heavy-ion collisions. The photon+jet measurements from CMS and ALTAS provide direct and less biased quantitative measures of jet energy loss in the medium, which give a deceasing jet-to-photon momentum imbalance ratio ($x_{j\gamma}$) from peripheral to central centrality bin in Pb+Pb collisions at LHC energy~\cite{Chatrchyan:2012gt, ATLAS:2012cna}. Some theoretical attempts have been made to understand it. Vitev and Zhang evaluated the transverse momentum imbalance of photon+jet is induced by the dissipation of parton shower energy due to strong final-state interactions~\cite{Dai:2012am}. Qin found that photon-tagged jet has a sensitivity on production position of $\gamma$-jet and propose it as a tomographic tool for studying jet quenching in heavy-ion collisions~\cite{Qin:2012gp}. In this Letter, a detail tomographic analysis with photon+jet are performed in Pb+Pb collisions at $\sqrt{s_{_{\rm NN}}}$ = 2.76 TeV within a multi-phase transport (AMPT) model with string meting scenario. The large imbalance of photon+jet can be produced by strong interactions between jet parton shower and partonic medium. Because the momentum imbalance is sensitive to both  production position and passing direction of $\gamma$-jet, it makes it possible to use photon+jet as a probe to do a detailed tomography on the partonic matter created in high energy heavy-ion collisions.

The Letter is organized as follows. In Section \ref{sec:model}, I give a brief description of AMPT model and jet physics inside, and introduce the analysis method for $\gamma$-jet reconstruction. Results on $\gamma$-jet imbalance are presented in Section \ref{sec:resul1} and then a possible detailed tomography with $\gamma$-jet is discussed in Section \ref{sec:resul2}. Finally a summary is given in Section \ref{sec:concl}.

\section{Model introduction and analysis method}
\label{sec:model}

The AMPT model with string meting scenario~\cite{Lin:2004en},  which has shown many good descriptions to some experimental observables~\cite{Lin:2004en,Chen:2006ub, Zhang:2005ni, Ma:2011uma, Ma:2010dv}, is implemented in this work. The AMPT model includes four main stages of high energy heavy-ion collisions: the initial condition, parton cascade, hadronization, and hadronic rescatterings. The initial condition, which includes the spatial and momentum distributions of minijet partons and soft string excitations, is obtained from HIJING model~\cite{Wang:1991hta,Gyulassy:1994ew}. Next it starts the parton evolution with a quark and anti-quark plasma from the melting of strings. The scatterings among these quarks are treated by using the Zhang's Parton Cascade (ZPC) model~\cite{Zhang:1997ej} with the differential elastic scattering cross section
\begin{equation}\label{cross}
\frac{d\sigma}{dt} \approx \frac{9\pi\alpha_s^2}{2(t-\mu^2)^2}.
\end{equation}
In the above, $t$ is the standard Mandelstam variable for four momentum transfer, $\alpha_s$ is the strong coupling constant, and $\mu$ is the screening mass in the partonic matter. It recombines partons via a simple coalescence model to produce hadrons when the partons freeze out. Dynamics of the subsequent hadronic matter is then described by ART model~\cite{Li:1995pra}.  In this work, the AMPT model with the newly fitted parameters for LHC energy~\cite{Xu:2011fi} is used to simulate Pb+Pb collisions at $\sqrt{s_{_{\rm NN}}}$ = 2.76 TeV. Two sets of partonic interaction cross sections, 0 and 1.5 mb, are applied to simulate two different physical scenarios for hadronic interactions only and parton + hadronic interactions, respectively.

In order to study the energy loss behaviors of $\gamma$-jet, a $\gamma$-jet of $p_{T}^{\gamma}\sim$ 60 GeV/c with known initial production position and direction is triggered with the jet triggering technique in HIJING,  since the production cross section of $\gamma$-jet is quite small especially for large transverse momentum. Three hard $\gamma$-jet production processes with high virtualities are additionally taken into account in the initial condition of the AMPT model, including $q+\bar{q}\rightarrow g+\gamma$, $q+\bar{q} \rightarrow \gamma+\gamma$ and $q+g \rightarrow  q+\gamma$~\cite{Sjostrand:1993yb}. For these prompt photons from $\gamma$-jet production, their birth information is kept for the analysis of $\gamma$-jet imbalance, since they only participate in electromagnetic interactions.  On the other hand, the high-$p_{T}$ primary partons pullulate to jet showers full of lower virtuality partons through initial- and final- state QCD radiations. In the string meting scenario of AMPT model, the jet parton showers are fragmented into hadrons with the LUND fragmentation, built in the PYTHIA routine~\cite{Sjostrand:1993yb}, and then these hadrons are converted into on-shell quarks and anti-quarks according to the flavor and spin structures of their valence quarks. In a sense, the melting scenario for jets, which bears some analogy to the medium-induced subsequent radiations,  but happens before jet-medium interactions in the logical structure of the AMPT model. After the melting process, not only a quark and anti-quark plasma is formed, but also jet quark shower is built up, therefore the initial configuration between $\gamma$-jet and the medium is ready to interact. In the following, the ZPC model automatically simulates all possible elastic partonic interactions among medium partons, between jet shower partons and medium partons, and among jet shower partons, but without the considerations of inelastic interactions or further radiations at present. Two sets of partonic interaction cross sections, 0 or 1.5 mb, are used to turn off or on the process of parton cascade in this study. When the partons freeze out, they are recombined into medium hadrons and jet shower hadrons via a simple coalescence model by combining the nearest partons into mesons and baryons. The final-state hadronic rescatterings including the interactions between jet shower hadrons and hadronic medium can be described by ART model~\cite{Li:1995pra}.

The kinetic cuts for the analysis on $\gamma$-jet transverse momentum imbalance are chosen as CMS experiment did~\cite{Chatrchyan:2012gt}. The transverse momentum of photon is required to be larger than 60 GeV/c ($p_{T}^{\gamma} >$ 60 GeV/c) and its pseudorapidity is within a mid-rapidity gap of 1.44 ($|\eta^{\gamma}|<1.44$). An anti-$k_{t}$ algorithm from the standard Fastjet package is made use of to reconstruct the full jet~\cite{Cacciari:2011ma}. Jet cone size is set to be 0.3 ($R$=0.3) , $p_{T}$ of jet is larger than 30 GeV/c ($p_{T}^{jet} >$ 30 GeV/c) and pseudorapidity of jet is within a mid-rapidity range of $|\eta^{jet}|<1.6$. A pseudorapidity strip of width $\Delta\eta$=1.0 centered on the jet position, with two highest-energy jets excluded, is used to estimate the background (``average energy per jet area"), which is subtracted from the reconstructed jet energy in Pb+Pb collisions. Both jet energy scale and jet efficiency corrections, which are obtained by embedding triggered p+p into non-triggered Pb+Pb events, have been applied for each jet. The $\gamma$-jet-triggered events are weighted with the experimental measured prompt photon $p_{T}$ spectra eventually~\cite{Grabowska-Bold:2012gza,Chatrchyan:2012vq}. 

\section{$\gamma$-jet transverse momentum imbalance}
\label{sec:resul1}

\begin{figure}
\includegraphics[scale=0.45]{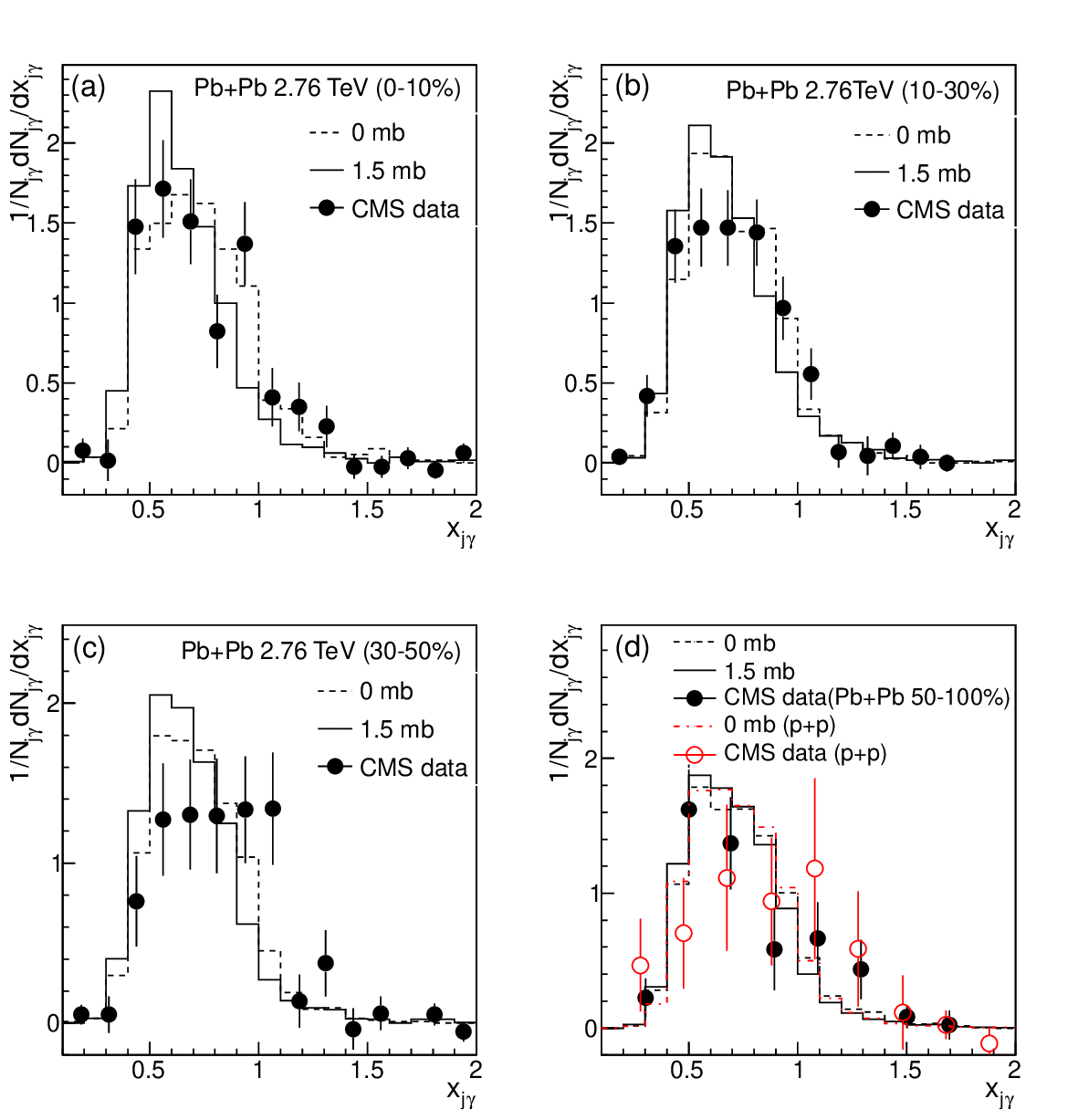}
\caption{(Color online) The distributions of imbalance ratio $x_{j\gamma}$=$p_{T}^{jet}$/$p_{T}^{\gamma}$ between the photon ($p_{T}^{\gamma} >$ 60 GeV/c) and jet ($p_{T}^{jet} >$ 30 GeV/c, $\Delta\phi_{j\gamma} > 7\pi/8$ ) after background subtraction for four centrality bins in Pb+Pb and p+p collisions, where the solid (1.5 mb) and dash (0 mb) histograms represent  the AMPT results with partonic+hadronic and hadronic interactions only respectively, while the circles represent the data from CMS experiment~\cite{Chatrchyan:2012gt}.}
 \label{fig-xjrdis}
\end{figure}

\begin{figure}
\includegraphics[scale=0.45]{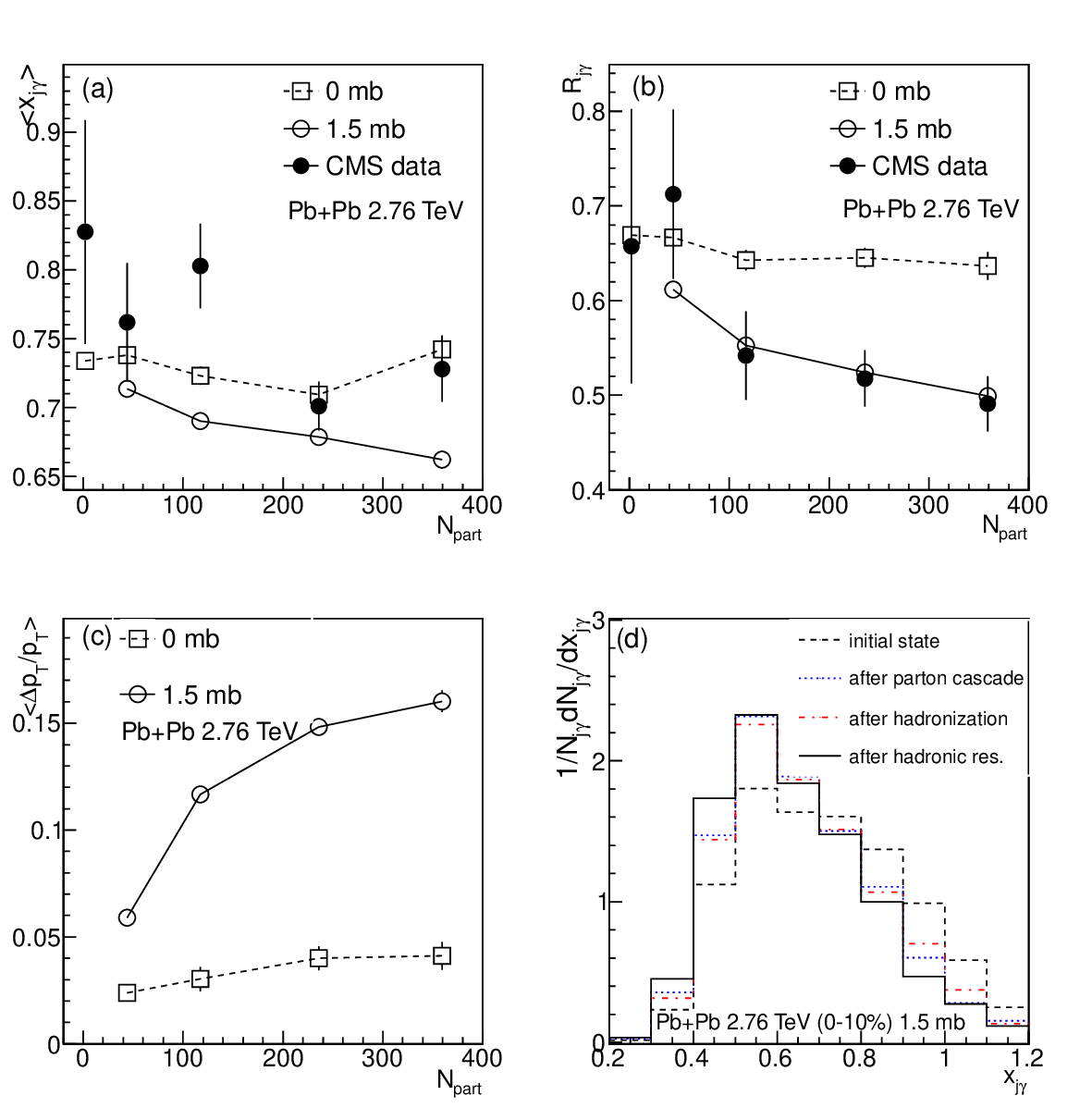}
\caption{(Color online) (a) Average ratio $\left\langle x_{j\gamma} \right\rangle$ as functions of $N_{part}$. (b)  Average ratio of photon with an associated jet above 30 GeV/c, $R_{j\gamma}$, as functions of $N_{part}$.  (c) Average energy loss fraction of jet, $\left\langle\Delta p_{T}/p_{T} \right\rangle$, as functions of $N_{part}$. (d) The distributions of imbalance ratio $x_{j\gamma}$ at or after different evolution stages in most central Pb+Pb collisions (1.5 mb). 
}
 \label{fig-meanandR}
\end{figure}

The transverse momentum imbalance is defined as the ratio of $x_{j\gamma}$=$p_{T}^{jet}$/$p_{T}^{\gamma}$ to study jet energy loss mechanism~\cite{Chatrchyan:2012gt, ATLAS:2012cna}. Figure~\ref{fig-xjrdis} (a)-(d) show the imbalance ratio distributions for four centrality bins in Pb+Pb collisions and p+p collisions at $\sqrt{s_{_{\rm NN}}}$ = 2.76 TeV.  The corresponding averaged values of imbalance ratio $\left\langle x_{j\gamma} \right\rangle$ as functions of number of participant nucleons $N_{part}$ are presented in Figure~\ref{fig-meanandR} (a). The AMPT results with both partonic and hadronic interactions (i.e. 1.5 mb) give a little smaller $x_{j\gamma}$ and $\left\langle x_{j\gamma} \right\rangle$ than those with hadronic interactions only (i.e. 0 mb) and experimental data.  On the other hand, Figure~\ref{fig-meanandR} (b) shows that only the AMPT result with both partonic and hadronic interactions can well reproduce the fraction $R_{j\gamma}$ of photons that have an associated jet with $p_{T}^{jet} >$ 30 GeV/c. These results basically support the picture of jet quenching in partonic matter at LHC. To quantitatively learn how much jet loses its energy in partonic or hadronic matter, the averaged energy loss fractions of jet, $\left\langle\Delta p_{T}/p_{T} \right\rangle$=$\left\langle (p_{T}^{jet,initial}-p_{T}^{jet,final})/p_{T}^{jet,initial} \right\rangle$, are shown for the four centrality bins in Figure~\ref{fig-meanandR} (c). Jet loses its energy from by $\sim$ 15\% in central collisions down to by $\sim$ 5\% in peripheral collisions due to decreasing of partonic interactions. However hadronic interactions with vanished partonic interactions only can give much smaller energy loss fraction around 4\%-2\%. It indicates that the strong interactions between jet parton shower and partonic matter can produce a larger momentum asymmetry than the interactions between jet hadron shower and hadronic matter, especially for more central collisions.

Since heavy-ion collisions actually is a dynamical evolution which consists of many important stages, it is very essential to see the effect separately from these stages on the imbalance.  Figure~\ref{fig-meanandR} (d) gives the distributions of imbalance ratio $x_{j\gamma}$ at or after different evolution stages for most central Pb+Pb collisions from AMPT simulations with both partonic and hadronic interactions (1.5 mb). It is found that jet indeed mainly lose its energy in the parton cascade process, because jet shower partons strongly interact with medium partons and lose energy into the partonic medium through frequent elastic parton interactions. The following hadronization from a jet parton shower to a jet hadron shower and hadronic rescattering between the jet hadron shower and the hadronic medium do not change the $x_{j\gamma}$ distribution much more. Therefore, photon+jet measurements basically can reflect the information about the interactions between jets and partonic matter.

\section{$\gamma$-jet detailed tomography}
\label{sec:resul2}

\begin{figure}
\includegraphics[scale=0.4]{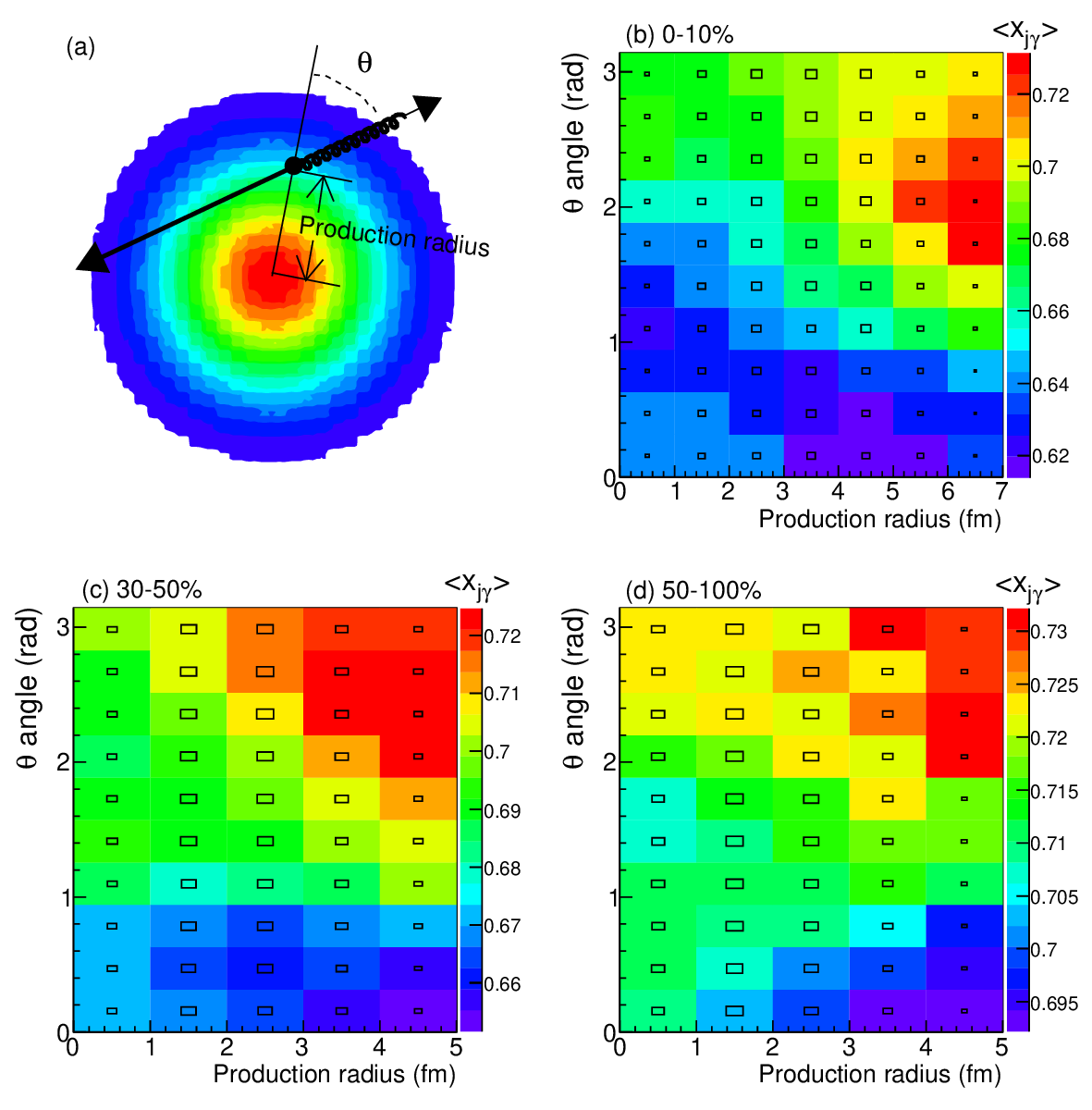}
\caption{(Color online) (a) Illustration of the production of $\gamma$-jet which passes through the partonic medium in a central Pb+Pb collision. See text for more detailed descriptions. (b)-(d) The production radius and direction dependences of averaged imbalance ratio, $\left\langle x_{j\gamma} \right\rangle$(r, $\theta$) for three selected centrality bins in Pb+Pb collisions from AMPT simulations (1.5 mb), where the color of cell denotes the $\left\langle x_{j\gamma} \right\rangle$ and the size of box in cell represents the production possibility for $\gamma$-jet with r and $\theta$.
}
 \label{fig-Rthetadis}
\end{figure}

\begin{figure}
\includegraphics[scale=0.45]{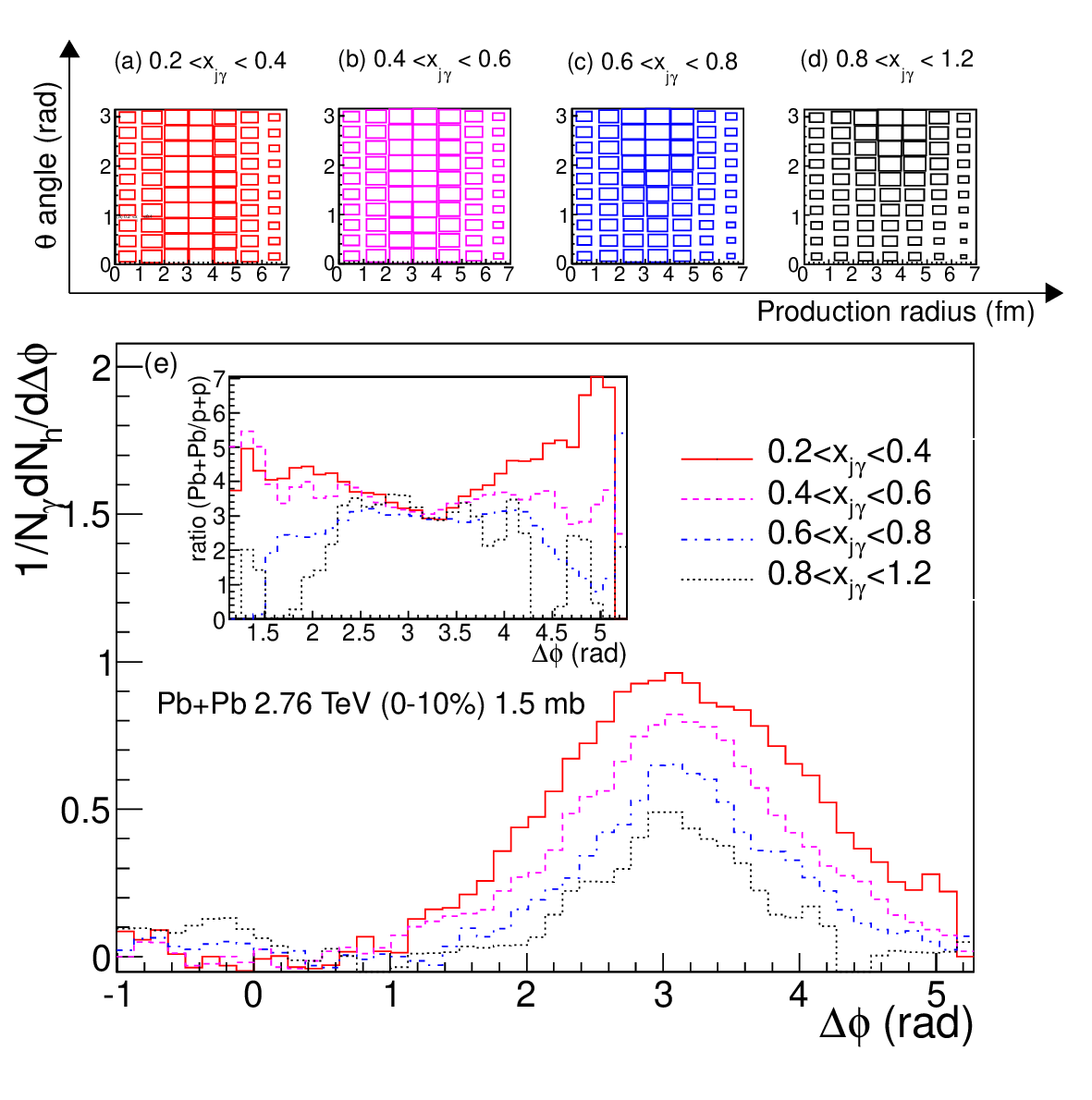}
\caption{(Color online)  (a)-(d) The possibility distributions of measured photon+jet events in $r$-$\theta$ plane with different $x_{j\gamma}$ selections for most central centrality bin (0-10\%) in Pb+Pb collisions. (e) The AMPT results on associated hadron ($p_{T} <$ 1 GeV/c, $|\eta| <$ 2) azimuthal correlations with a triggered photon ($p_{T}^{\gamma} >$ 60 GeV/c) for the most central Pb+Pb events with different $x_{j\gamma}$ ranges, where the inserted panel shows the corresponding away-side $\gamma$-hadron correlation ratios of most central Pb+Pb to p+p collisions. }
 \label{fig-gammahadron}
\end{figure}

However, all above are the inclusive results from the average of all cases for $\gamma$-jets with any possible production position and direction, it is difficult to learn the details about how jets probe the medium and how the medium responses to jets. As Figure~\ref{fig-Rthetadis} (a) illustrates, a pair of photon and jet can be produced at a position (the black dot), where is at a distance of production radius $r$ to the center, in a central Pb+Pb collision. The direction of photon can be represented by a $\theta$ angle which is the angle between the passing direction of photon and the vector from the center to production point. The current measured imbalance between photon and jet actually is an inclusive case for all possible production radii and $\theta$ angles. Therefore it is helpful to do a differential jet tomographic analysis for understanding the properties of new form of matter in more details. Figure~\ref{fig-Rthetadis} (b)-(d) shows the production radius and direction dependence of imbalance ratio of $\gamma$-jets, i.e. the averaged value $\left\langle x_{j\gamma} \right\rangle$ at (r, $\theta$), for the three central centrality bins in Pb+Pb collisions from AMPT simulations with both partonic and hadronic interactions (1.5 mb), where the color of cell denotes the $\left\langle x_{j\gamma} \right\rangle$ and the size of box in cell represents the production possibility for $\gamma$-jet with $r$ and $\theta$. Note that $\gamma$-jets actually are produced anisotropically due to the asymmetric geometries for non-central collisions, but the representation of ($r$, $\theta$) is taken for simplicity.   It is interesting that the $\left\langle x_{j\gamma} \right\rangle$ is sensitive to both production position and direction of $\gamma$-jet. These $\gamma$-jet production configurations basically can be divided into three typical cases. (I) Pouch-through jet case: The $\gamma$-jets with large production radii and small $\theta$ angles tend to have small $x_{j\gamma}$ values, because jet parton showers punch long paths through the partonic medium and lose much energy.  (II) Escaped-jet case: The $\gamma$-jets with large production radii and large $\theta$ angles prefer to keep the original initial momentum balances (i.e. large $x_{j\gamma}$), since most of jet shower patrons directly escape out of the partonic medium without any or with few interactions in very short lengths. (III) Tangential-jet case: The $\gamma$-jets with large production radii and $\theta$ angles $\sim \pi/2$, tangentially passing the medium, have middle $x_{j\gamma}$ values between the two above cases. The two-dimensional dependences of $\left\langle x_{j\gamma} \right\rangle$(r, $\theta$) indicate that final $x_{j\gamma}$ is sensitive to the initial $\gamma$-jet production information to a certain degree. However it should be pointed out that the initial geometry sensitivity of $x_{j\gamma}$ is only based on the AMPT calculations, which possibly is weaken or even lost for different models~\cite{Renk:2009gn}. Based on these dependences , the different $\gamma$-jet production configurations can be classified by selecting $x_{j\gamma}$ range.  For instance, Figure~\ref{fig-gammahadron} (a)-(d) show the possibility distributions of measured photon+jet events in $r$-$\theta$ plane with different $x_{j\gamma}$ range selections.  The $\gamma$-jet events with smallest $x_{j\gamma}$ values, e.g. (a) $0.2 < x_{j\gamma} < 0.4$, have more punch-through jet components. The $\gamma$-jet events with highest $x_{j\gamma}$ values, e.g. (d) $0.8 < x_{j\gamma} < 1.2$, prefer  to be the escaped-jet case. The $\gamma$-jet events with the modest $x_{j\gamma}$ values, e.g. (b) $0.4 < x_{j\gamma} < 0.6$, correspond to with more tangential-jet components.

The medium could be excited by jet shower propagation inside a quark-gluon plasma~\cite{CasalderreySolana:2004qm, Stoecker:2004qu}. However the photon+jet measurement misses the main part of medium excitations which are mostly out of jet cone size of $R$. To study the medium response, $\gamma$-hadron azimuthal correlation, which includes all particles correlated with $\gamma$, was proposed as a golden probe because the flow background subtraction becomes trivial for $\gamma$-triggered correlation in comparison with jet-triggered one~\cite{Ma:2010dv, Li:2010ts}. Actually $\gamma$-hadron azimuthal correlation in combination with photon+jet can work more efficiently. Figure~\ref{fig-gammahadron} (e) shows the $\gamma$-hadron azimuthal correlation under the different $x_{j\gamma}$ selections, where the more enhanced away-side peak is observed with the decreasing of $x_{j\gamma}$. The inserted plot of Figure~\ref{fig-gammahadron} (e) displays the corresponding ratios of $\gamma$-hadron correlations for away-side between central Pb+Pb and p+p collisions for different $x_{j\gamma}$ ranges. It is pronounced to observe the large enhancement of two side peaks in small $x_{j\gamma}$ classes of Pb+Pb events, which indicates the deflection of both Mach-cone like medium excitation and jet shower by radial flow~\cite{Li:2010ts} is more likely formed for the punch-through configurations of $\gamma$-jet in central Pb+Pb collisions. It is proposed to measure $\gamma$-hadron correlations associated with different $x_{j\gamma}$ conditions to zoom in on the medium responses experimentally.

\section{Summary}
\label{sec:concl}

In summary, the transverse momentum imbalance between prompt photon and jet is analyzed in the framework of a multi-phase transport model. The large transverse momentum imbalance is produced by strong partonic interactions between jet parton showers and partonic medium. The hadronization and final-state hadronic interactions have little effect on the final measured imbalance. Within the AMPT simulations, the imbalance ratio $x_{j\gamma}$ shows sensitivity to both production position and the passing direction of $\gamma$-jet, which possesses a potential prospect in a detailed $\gamma$-jet tomography on the new form of matter at RHIC and LHC. $\gamma$-hadron azimuthal correlation with the help of $\gamma$-jet imbalance is also proposed as a probe to see medium responses to different $\gamma$-jet production configurations.

\section*{Acknowledgements}
I thank X. -N. Wang, G. Y. Qin, F. Q. Wang, Y. X. Mao, Y. G. Ma, J. Xu and Q. Y. Shou for helpful discussions. This work was supported by the NSFC of China under Projects Nos. 11175232, 11035009, 11105207, U1232206, 11220101005, the Knowledge Innovation Program of CAS under Grant No. KJCX2-EW-N01, Youth Innovation Promotion Association, CAS, the Project-sponsored by SRF for ROCS, SEM and CCNU-QLPL Innovation Fund (QLPL2011P01).


\end{document}